# High Purity Orbital Angular Momentum of Light

ANDREW SONTAG,[1] MEHMET A. NOYAN,[1,2] AND JAMES M. KIKKAWA[1,*]

[1]*Department of Physics & Astronomy, The University of Pennsylvania, Philadelphia, PA 19104, USA*
[2]*Current Address: Materials Science and Technology Division, Naval Research Laboratory, Washington, DC 20375, USA*
*\*kikkawa@physics.upenn.edu*

**Abstract:** We present a novel technique for generating beams of light carrying orbital angular momentum (OAM) that increases mode purity and decreases singularity splitting by orders of magnitude. This technique also works to control and mitigate beam divergence within propagation distances less than the Rayleigh length. Additionally, we analyze a tunable parameter of this technique that can change the ratio of beam purity to power to fit desired specifications. Beam generation via this technique is achievable using only phase-modulating optical elements, which reduces experimental complexity and beam energy loss.

## 1. Introduction

Light carrying orbital angular momentum (OAM) or 'vortex light' [1-4] has come under increased study for applications in optical tweezers and micromotors [5,6], improving astrophysical observation [7], increasing communication channel capacity [8-10], and facilitating the next generation of quantum computation [11,12]. Traditional methods of generating vortex beams use spiral phase plates [13], q-plates [14,15], metasurfaces [16], and holographic amplitude or phase gratings [6]. All of these methods, as they have been traditionally implemented, have a critical imperfection that yields low purity modes with anomalies in amplitude and phase. Prominent among these anomalies is singularity splitting, a phenomenon by which a single high order phase singularity will split into a number of singularities of order ±1. This effect is well documented [17-25] and numerous attempts have been made to correct for it [26-34]. However, the aforementioned methods suffer from a number of issues, including: requirement of a nonlinear propagating medium [26,27], requirement of simultaneous amplitude and phase modulation [28,29], iterative measurement-based methodology [30], and low maximal mode purity [30-33]. Mode purity is increasingly important and even crucial for certain OAM proposals. For example, mode impurity is the largest obstacle to reducing crosstalk in multiplexed OAM communication systems [35,36], and is also the limiting factor for increasing nanoscale microscopy resolution [37] and rotational sensitivity [38] beyond the diffraction limit. Mode purity protects propagation stability, which is vital in quantum computation to decrease data transfer error rates and lengthen decoherence times [39]. High purity OAM is additionally desirable for quantum material analysis [40,41], and is essential for proposals of ultra-relativistic electron acceleration where higher order radial modes limit electron beam collimation [42]. In this paper, we introduce a fundamental correction to OAM holographic gratings that reduces amplitude and phase anomalies by orders of magnitude yet requires minimal experimental complexity to implement.

## 2. Method

Traditional spiral phase plates/mirrors, q-plates, metasurfaces, and holographic amplitude and phase gratings strive to impart an azimuthal phase dependence $\exp(-im\theta)$, and thus a phase singularity of order $m$, to an input Gaussian beam. OAM modes created via these methods suffer from prominent singularity splitting for $|m| > 1$ upon propagation because they are superpositions of a number of Laguerre-Gaussian (LG) modes. LG modes of radial index $p$ and azimuthal index $m$ are solutions to the paraxial wave equation that have phase singularities of order $m$ and carry $m\hbar$ OAM per photon [1,2]. A superposition of some desired mode $LG_p^m$ and any $LG_p^0$ modes will cause the phase singularity of order $m$ to split into $|m|$ singularities, each of order $\text{sgn}(m)$ [17,18]. Other superpositions produce more complicated splitting patterns. When $p = 0$, the complex amplitude of $LG_p^m$ is proportional to $\exp\left(-\frac{r^2}{w^2}\right)\left(\frac{r}{w}\right)^{|m|} \exp(-im\theta)$ for beam waist $w$. The traditional generation methods create superpositions of multiple LG modes because the beams they produce approximate the form $\exp\left(-\frac{r^2}{w^2}\right)\exp(-im\theta)$, which is missing the $\left(\frac{r}{w}\right)^{|m|}$ term of a pure LG mode.

One possible improvement, then, is to create a hybrid amplitude-phase method by taking any of the traditional methods and applying an additional transmittance mask proportional to $\left(\frac{r}{r_{\max}}\right)^{|m|}$, for some minimal $r_{\max}$ such that the transmittance is $\leq 1$ everywhere. However, for the sake of practicality, we seek to design an optical element that improves mode purity using only phase modulation, so that our method may be easily implemented with a single two-dimensional spatial light modulator (2D SLM). We find that we can improve mode purity by taking the retardance distribution of the traditional holographic phase grating, given by $\frac{\lambda}{4}(1 + \cos(ax + m\theta))$ [6], multiplying that traditional distribution by $\left(\frac{r}{r_{\max}}\right)^{|m|}$ and using an additional term to correct for the Rayleigh range curvature.

Due to the spherical spreading of an LG mode over long propagation distances, the complex amplitude of every LG mode carries a factor of $\exp(-ikr^2z/(2(z^2 + z_R^2)))$, where $z_R = kw^2/2$ is the Rayleigh range [1,3]. This factor causes the lines of constant phase in a planar cross section of the beam to be curved near the center of the beam at every propagation distance except $z = 0$, which is the focal point of the beam. This Rayleigh range curvature can be corrected by adding an additional term to the grating retardance of the form $-r^2z/(2(z^2 + z_R^2))$. The combination of the above $\left(\frac{r}{r_{\max}}\right)^{|m|}$ retardance scaling and the Rayleigh range curvature correction results in what we will refer to as the "high purity" grating, with retardance distribution equal to:

$$R(\vec{r}) = C + \frac{\lambda}{4}(1 + \cos(ax + m\theta))\left(\frac{r}{r_{\max}}\right)^{|m|} - \frac{r^2 z}{2(z^2 + z_R^2)}$$

where $C$ is a constant simply used to ensure compatibility with the SLM retardance window. The difference between the traditional and high purity (with $r_{\max} = 1.5w$) grating patterns is shown in Figure 1 for an $m = +1$ grating, although the same technique can successfully be applied to a grating of any $m$ value.

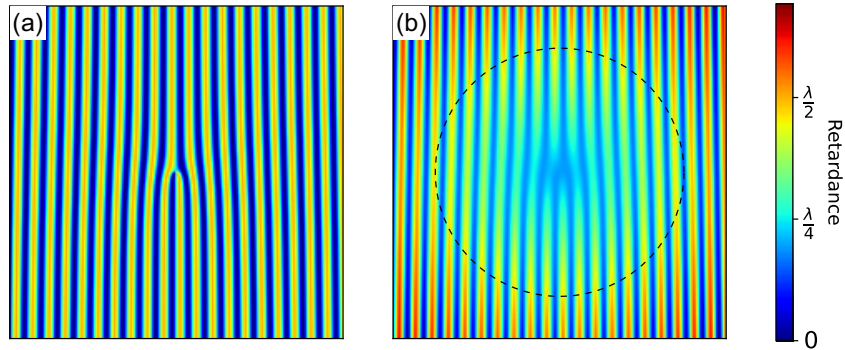

Fig. 1. Central sections of the retardance distributions for traditional (a) and high purity (b) holographic $m = +1$ phase gratings. The dashed line indicates the waist of the incident beam.

To analyze the effects of the grating correction, we use a vectorial Huygens-Fresnel simulation to compare the first-order diffracted peaks produced at a distance of 5 m when a continuous wave Gaussian beam of wavelength 800 nm and various different beam waists is incident upon a traditional grating versus a high purity grating. Details of the simulation method are given in the Supporting Information. An example of one such simulated output is given in Figure 2, for $m = +3$ traditional and high purity gratings. The improvement of the phase distribution caused by the correction is clear from the phase plots, and the decrease in singularity splitting radius (or, equivalently, increase in available propagation distance before splitting occurs) can be quantitatively seen in the Poynting vector skew angle plots. For a perfectly pure LG mode, the Poynting vector skew angle will follow the black curve [43]. For an impure mode, the radius at which the skew angle begins deviating significantly from its ideal value provides a robust estimate of the splitting radius. We find that the correction causes similar improvements for all other $m$ values with $|m| > 1$, with an increase in the magnitude of $m$ bringing about larger splitting radii in both the traditional and high purity cases. For $m = \pm 1$, the singularity cannot split, so we must use a different measure of beam improvement.

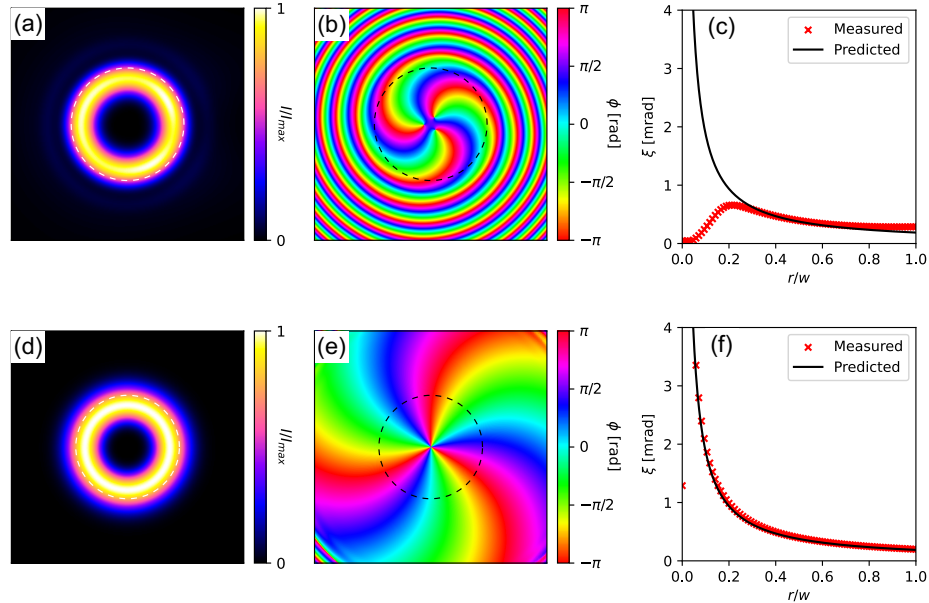

Fig. 2. Intensity (a,d), phase (b,e), and Poynting vector skew angle (c,f) compared for first-order diffracted peaks of $m = +3$ traditional (a,b,c) versus high purity (d,e,f) gratings. Dashed lines indicate beam waists.

For LG modes with all nonzero $m$ values, a rough visual indicator of beam impurity is the appearance of unexpected ripples in the radial intensity profile. These ripples are visible just outside the main radial intensity peak in Figure 2, and are made more visible by looking at the fractional intensity error (relative to the ideal $\text{LG}_0^m$ mode) in Figure 3. The intensity profile for traditional $m = +3$ gratings differs from the ideal case by up to 30% in ripples away from the center (Fig. 3b). The corrected profile (Fig. 3d) is improved by more than two orders of magnitude. This result extends to all nonzero $m$ values, again with the deviation from the expected intensity generally growing in both cases as $|m|$ increases.

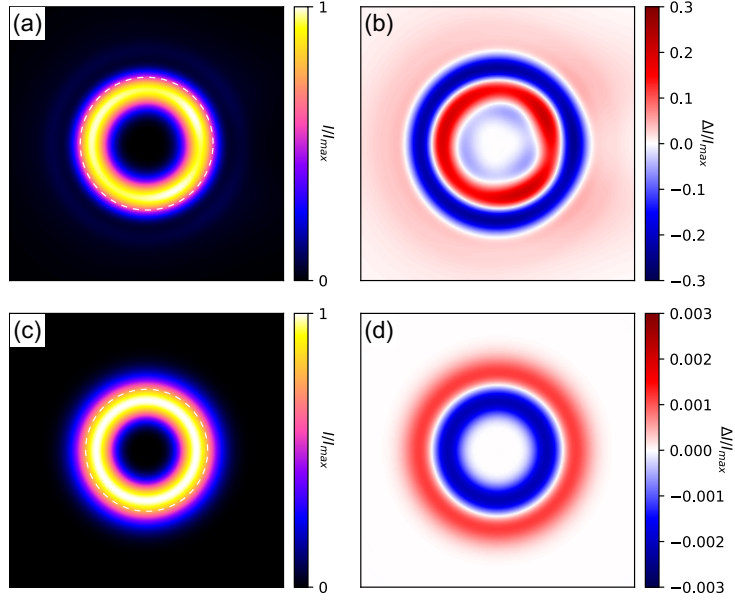

Fig. 3. Intensity (a,c) and error (b,d) compared for first-order diffracted peaks of $m = +3$ traditional (a,b) versus high purity (c,d) gratings. Note the factor of 100 contrasting the two error scales. Dashed lines indicate beam waists.

To further quantify beam purity, we note that the LG modes with integer indices $p$ and $m$ ($p \geq 0$) form a complete and orthogonal set of functions [44], so we can take any intensity and phase distribution and decompose it into a sum over the LG modes. We define the complex amplitude of an output mode to be

$$\psi = \sqrt{|\vec{S}|} \exp(i\phi)$$

for Poynting vector $\vec{S}$ and phase $\phi$, such that the intensity is given by $|\psi|^2$. Using this definition, we write:

$$\psi = \sum_{m'\in\mathbb{Z}} \sum_{p'=0}^{\infty} c_{m',p'} \text{LG}_{p'}^{m'}$$

where the modes $\text{LG}_{p'}^{m'}$ are scaled such that they are orthonormal over the beam cross section. We then interpret $|c_{m',p'}|^2$ to be the intensity carried by the $\text{LG}_{p'}^{m'}$ mode within the beam. We normalize $\psi$ such that the sum over all $|c_{m',p'}|^2$ is unity, and then analyze the intensity distribution of a given beam in $(m', p')$ space to quantitatively assess its purity. One example of such an analysis is shown in Figure 4. Across all nonzero $m$ values we see similar trends, with the output beam for the traditional grating (Fig. 4a) being relatively smooth in $p'$ space and having a strong $m' = 0$ contamination, while the high purity beam (Fig. 4b) is sharply concentrated in the desired $m' = m$ value and has a steep exponential decline with increasing $p'$.

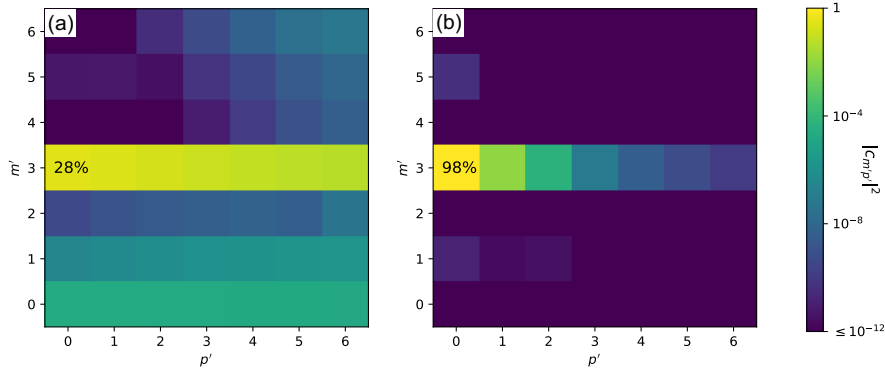

Fig. 4. $LG_{p'}^{m'}$ coefficients $|c_{m',p'}|^2$ for first-order diffracted peaks of $m = +3$ traditional (a) and high purity (b) gratings. The percentage shown on each plot is $|c_{3,0}|^2$, the intensity fraction in the desired mode.

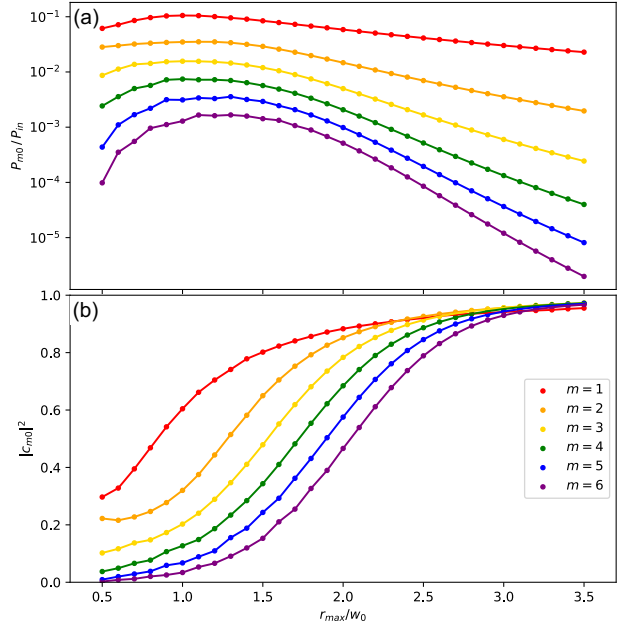

Fig. 5. Plots of power in desired mode relative to power in incident Gaussian (a) and relative to total power in first-order diffracted peak (b) versus SLM retardance scaling parameter $r_{\text{max}}$, for $m = +1$ through $+6$ high purity gratings.

All of the previous data was generated with a retardance scaling parameter $r_{\text{max}} = 7.8w$, because it kept the retardance on our entire SLM between 0 and $\lambda/2$. But this restriction is not necessary. Varying the number $r_{\text{max}}/w$ allows us to control the power and purity of our output beam. We show the effects of varying $r_{\text{max}}/w$ for beams with $m = +1$ through $+6$ in Figure 5. Note that the ratio of power in desired mode to total power in first-order diffracted peak in panel (b) is equivalent to the purity measure $|c_{m,0}|^2$ as defined for Figure 4. The general trend is that higher values of $r_{\text{max}}/w$ yield purer beams with less power. The power drop is to be expected, as increasing $r_{\text{max}}$ decreases the retardance grating contrast in the region of the SLM where the peak incident intensity falls, and so increases the percentage of power diverted to the $0^{\text{th}}$-order diffracted peak. The purity grows with $r_{\text{max}}$ because, as the retardance difference between two equidistant paths increases through $\lambda/2$, the progressive deepening of their interference null is reversed. In regions of the SLM for which $r > r_{max}$, the full retardance span exceeds $\lambda/2$ and therefore no longer represents the intended condition of maximal destructive interference between equidistant paths. Contributions to the output mode from $r > r_{max}$ no longer create destructive interference in the correct locations, creating a noisier output mode. We observe that beams with lower $m$ values attain higher purities faster as $r_{max}/w$ increases, but that all $m$ values see purities approaching unity.

To highlight the effect of changing $r_{max}$, in Figure 6 we show the first-order diffracted peak of the high purity $m = +3$ grating with three different values of $r_{max}$. First, we have $r_{max}/w = 0.5$, showcasing how the mode quality decays significantly with small $r_{max}/w$. Next, we show $r_{max}/w = 1.2$, which is the value of $r_{max}/w$ that maximizes the power in the $LG_0^3$ mode. Some singularity splitting is observed, but it is still a significant improvement over the traditional grating output splitting shown in Figure 2. Finally, we show $r_{max}/w = 3$, with no visible singularity splitting, as a high purity example. This analysis shows how $r_{max}$ can be tuned to accommodate a wide variety of power and/or purity requirements.

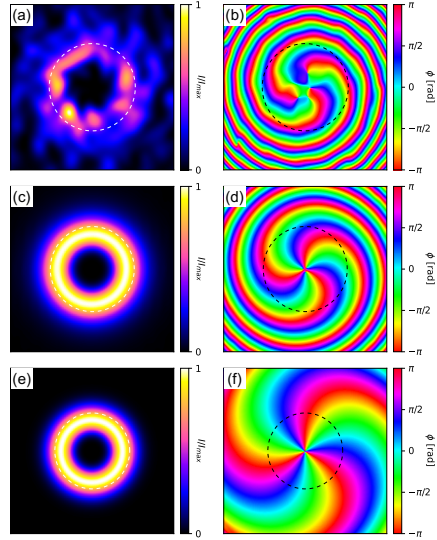

Fig. 6. Intensity (a,c,e) and phase (b,d,f) of the first-order diffracted peaks of high purity $m = +3$ gratings with retardance scaling parameters $r_{max} = 0.5w$ (a,b), $r_{max} = 1.2w$ (c,d), and $r_{max} = 3w$ (e,f). Dashed lines indicate beam waists.

We observe that, for the high purity grating, the Rayleigh range curvature correction term seems to play a more vital role in preventing singularity splitting than one might expect. Even though the phase factor in the LG mode complex amplitude causing the Rayleigh range curvature cannot alone split the singularity, we find that removing the added Rayleigh range curvature correction term from the high purity grating retardance yields beams with split singularities and qualitatively similar purity to the traditional grating. However, adding the Rayleigh range curvature correction term by itself to a traditional grating yields almost no change in output. Thus, the high purity grating requires both the Rayleigh range correction and the radial phase factor. A comparison of traditional and high purity gratings, with and without the Rayleigh range curvature correction term, is shown in Figure 7.

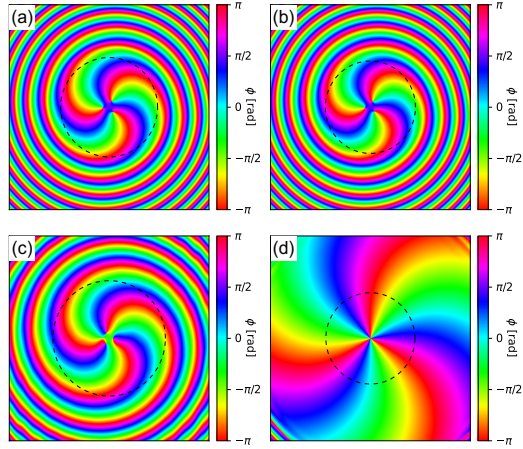

Fig. 7. Phase distributions of the first-order diffracted peaks of $m = +3$ traditional (a,b) and high purity (c,d) gratings, with the Rayleigh range curvature correction term excluded (a,c) and included (b,d). Dashed lines indicate beam waists.

Interestingly, adding the $\left(\frac{r}{r_{\max}}\right)^{|m|}$ scaling factor to the traditional grating as a *transmission* term rather than a retardance term, as alluded to in the motivation for its conception, removes this strict dependency on the Rayleigh range curvature correction term. The SLM composed of a traditional grating along with an $\left(\frac{r}{r_{\max}}\right)^{|m|}$ transmission term and the Rayleigh range curvature correction term achieves the same purification effects as does the high purity grating; however, removing the Rayleigh range curvature correction term does not bring about singularity splitting, but merely extraneous curvature in the lines of constant phase within the cross section of the beam. Furthermore, we find that adding an $\left(\frac{r}{r_{\max}}\right)^{|m|}$ transmission term to both traditional spiral phase plates and traditional q-plates (as described in [13] and [14,15]) produces the same results as the combination retardance-transmission grating described above (though for these the desired mode is the $0^{th}$-order diffracted peak rather than the $1^{st}$), with the same relationship to the Rayleigh range curvature correction term. We include this digression to point out that, in the event that a Rayleigh range curvature correction term is somehow unachievable, or that a spiral phase plate or q-plate is specifically desirable over a holographic grating, adding an $\left(\frac{r}{r_{\max}}\right)^{|m|}$ transmission term will still function to prevent singularity splitting at the cost of a more complicated experimental setup.

## 3. Conclusion

We have presented a novel technique for generating OAM modes with high purity in order to avoid singularity splitting upon propagation. High purity gratings are achieved while only modulating the phase of an incident Gaussian beam. The correction technique is tunable to different power and purity requirements in order to optimize a variety of different situations. In particular, mode purity can be raised arbitrarily high in order to lengthen decoherence times to any given specification for a multi-level quantum computer. Modifying the OAM in a beam by circulating it through multiple consecutive gratings also compounds the purity losses accrued by traditional gratings, making correction even more vital for computations with many steps. The redirection of power into the $0^{th}$-order diffracted peak (rather than loss via partial transmission) leaves open the possibility of recycling unused beam power to mitigate the cost of high purity. Further fine-tuning may also be possible by complicating the retardance multiplier as in [31,45,46]. Although all of the above simulation and analysis has been performed for electromagnetic waves, OAM matter waves have also been generated using traditional gratings [47-49], and the high purity grating we present could be used for improving matter wave purity as well.

**Funding.** Filler text, to be replaced upon processing.

**Acknowledgments.** Simulations described here were initiated and received primary support under NSF DMR-1206270. The analysis of high purity corrections and gratings was completed with the assistance of NSF MRSEC DMR-1720530.

**Disclosures.** The authors declare no conflicts of interest.

**Data availability.** Data underlying the results presented in this paper are not publicly available at this time but may be obtained from the authors upon reasonable request.

**Supplemental document.** See Supplement 1 for supporting content.

## References

1. L. Allen, M. W. Beijersbergen, R. J. C. Spreeuw, and J. P. Woerdman, "Orbital angular momentum of light and the transformation of Laguerre-Gaussian laser modes," Phys. Rev. A **45**, 8185 (1992).
2. M. Padgett, J. Courtial, and L. Allen, "Light's orbital angular momentum," Phys. Today **57**, 35 (2004).
3. S. M. Barnett, L. Allen, R. P. Cameron, C. R. Gilson, M. J. Padgett, F. C. Speirits, and A. M. Yao, "On the natures of the spin and orbital parts of optical angular momentum," J. Opt. **18**, 064004 (2016).
4. Y. Shen, X. Wang, Z. Xie, C. Min, X. Fu, Q. Liu, M. Gong, and X. Yuan, "Optical vortices 30 years on: OAM manipulation from topological charge to multiple singularities," Light Sci. Appl. **8**, (2019).
5. A.T. O'Neil, I. MacVicar, L. Allen, and M.J. Padgett, "Intrinsic and extrinsic nature of the orbital angular momentum of a light beam," Phys. Rev. Lett. **88**, 053601 (2002).
6. A. V. Carpentier, H. Michinel, J.R. Salgueiro, and D. Olivieri, "Making optical vortices with computer-generated holograms," Am. J. Phys. **76**, 916 (2008).
7. G. Berkhout and M. W. Beijersbergen, "Using a multipoint interferometer to measure the orbital angular momentum of light in astrophysics," Journal of Optics A: Pure and Applied Optics **11**, 9 (2009).
8. N. Bozinovic, Y. Yue, Y. Ren, M. Tur, P. Kristensen, H. Huang, A.E. Willner, and S. Ramachandran, "Terabit-Scale Orbital Angular Momentum Mode Division Multiplexing in Fibers," Science **340**, 1545 (2013).
9. C. Chen, G. Zhou, G. Zhou, M. Xu, Z. Hou, C. Xia, and J. Yuan, "A multi-orbital-angular-momentum multi-ring micro-structured fiber with ultra-high-density and low-level crosstalk," Opt. Commun. **368**, 27 (2016).
10. P. Z. Dashti, F. Alhassen, and H.P. Lee, "Observation of orbital angular momentum transfer between acoustic and optical vortices in optical fiber," Phys. Rev. Lett. **96**, 1 (2006).
11. J. C. Garcia-Escartin and P. Chamorro-Posada, "Quantum computer networks with the orbital angular momentum of light," Phys. Rev. A. **86**, 032334 (2012).
12. J. H. Lopes, W. C. Soares, B. Bernardo, D. P. Caetano, and A. Canabarro, "Linear optical CNOT gate with orbital angular momentum and polarization," Quant. Info. Proc. **18**, 256 (2019).
13. K. Sueda, G. Miyaji, N. Miyanaga, and M. Nakatsuka, "Laguerre-Gaussian beam generated with a multilevel spiral phase plate for high intensity laser pulses," Opt. Express **12**, 15, (2004).
14. L. Marrucci, C. Manzo, and D. Paparo, "Optical spin-to-orbital angular momentum conversion in inhomogeneous anisotropic media," Phys. Rev. Lett. **96**, 163905, (2006).
15. B. Piccirillo, V. D'Ambrosio, S. Slussarenko, L. Marrucci, and E. Santamato, "Photon spin-to-orbital angular momentum conversion via an electrically tunable q-plate," Appl. Phys. Lett. **97**, 241104, (2010).
16. R. Devlin, A. Ambrosio, N. Rubin, J. Mueller, and F. Capasso, "Arbitrary spin-to–orbital angular momentum conversion of light," Science, **358**, 6365, (2017).
17. I. V. Basistiy, V.Y. Bazhenov, M.S. Soskin, and M. V. Vasnetsov, "Optics of light beams with screw dislocations," Opt. Commun. **103**, 422 (1993).
18. F. Ricci, W. Loffler, and M. P. van Exter, "Instability of higher-order optical vortices analyzed with a multi-pinhole interferometer," Optica, **20**, 20, (2012).
19. M. R. Dennis, "Rows of optical vortices from elliptically perturbing a high-order beam," Opt. Lett. **31**, 1325 (2006).
20. I. Freund, "Critical point explosions in two-dimensional wave fields," Opt. Commun. **159**, 99 (1999).
21. Indebetouw, Guy. "Optical vortices and their propagation," J. Mod. Opt. **40**, 1 (1993).
22. B. Mao, Y. Liu, W. Chang, L. Chen, M. Feng, H. Guo, J. He, and Z. Wang, "Singularities splitting phenomenon for the superposition of hybrid orders structured lights and the corresponding interference discrimination method," Nanophotonics, **11**, 7, (2022).
23. Y. Zhang, Z. Wu, K. Yang, P. Li, F. Wen, and Y. Gu, "Splitting, generation, and annihilation of phase singularities in non-coaxial interference of Bessel–Gaussian beams," Phys. Scr. **96**, 2, (2021).
24. A. Mamaev, M. Saffman, and A. Zozulya, "Vortex Evolution and Bound Pair Formation in Anisotropic Nonlinear Optical Media," Phys. Rev. Lett. **77**, 22, (1996).

25. A. Mamaev, M. Saffman, and A. Zozulya, "Decay of High Order Optical Vortices in Anisotropic Nonlinear Optical Media," Phys. Rev. Lett. **78**, 11, (1997).
26. A. Desyatnikov, Y. Kivshar, and L. Torner, "Optical vortices and vortex solitons," Progress in Optics, **47**, (2005).
27. X. Gan, P. Zhang, S. Liu, Y. Zheng, J. Zhao, and Z. Chen, "Stabilization and breakup of optical vortices in presence of hybrid nonlinearity," Opt. Expr. **17**, 25, (2009).
28. T. Ando, N. Matsumoto, Y. Ohtake, Y. Takiguchi, and T. Inoue, "Structure of optical singularities in coaxial superpositions of Laguerre–Gaussian modes," J. Opt. Soc. A. **27**, 12, (2010).
29. S. Zheng, H. Hao, Y. Tang, and X. Ran, "High-purity orbital angular momentum vortex beam generator using an amplitude-and-phase metasurface," Opt. Lett. **46**, 23, (2021).
30. R. Neo, S.J. Tan, X. Zambrana-Puyalto, S. Leon-Saval, J. Bland-Hawthorn, and G. Molina-Terriza, "Correcting vortex splitting in higher order vortex beams," Opt. Expr. **22**, 9920 (2014).
31. T. Ando, Y. Ohtake, N. Matsumoto, T. Inoue, and N. Fukuchi, "Mode purities of Laguerre–Gaussian beams generated via complex-amplitude modulation using phase-only spatial light modulators," Opt. Lett. **34**, 1, (2009).
32. D. Gao and Y. Wang, "Generating high-purity orbital angular momentum vortex waves from Cassegrain meta-mirrors," Eur. Phys. J. Plus, **135**, 921, (2020).
33. A. Kumar, P. Vaity, and R. Singh, "Crafting the core asymmetry to lift the degeneracy of optical vortices," Opt. Expr. **19**, 7, (2011).
34. A. Kumar, P. Vaity, J. Bhatt, and R.P. Singh, "Stability of higher order optical vortices produced by spatial light modulators," J. Mod. Opt. **60**, 1696 (2013).
35. X. Heng, J. Gan, Z. Zhang, J. Li, M. Li, H. Zhao, Q. Qian, S. Xu, and Z. Yang, "All-fiber stable orbital angular momentum beam generation and propagation," Opt. Expr. **26**, 13, (2018).
36. S. Jiang, C. Chen, J. Ding, H. Zhang, and W. Chen, "Alleviating orbital-angular-momentum-mode dispersion using a reflective metasurface," Phys. Rev. Applied, **13**, 5, (2020).
37. J. Keller, A. Schonle, and S. Hell, "Efficient fluorescence inhibition patterns for RESOLFT microscopy," Opt. Expr. **15**, 6, (2007).
38. H. Shi and M. Bhattacharya, "Optomechanics based on angular momentum exchange between light and matter," J. Phys. B: At. Mol. Opt. Phys. **49**, 15, (2016).
39. J. Pan, C. Simon, C. Brukner, and A. Zeilinger, "Entanglement purification for quantum communication," Nature, **410**, 1067-1070, (2001).
40. A. Abdalla, S. Abdelmajid, and K. Kuacgor, "Topological spin textures of exciton-polaritons manipulating by spin-orbit coupling," Res. In Opt. **3**, (2021).
41. K. Sun, C. Qu, and C. Zhang, "Spin–orbital-angular-momentum coupling in Bose-Einstein condensates," Phys. Rev. A. **91**, 6, (2015).
42. D. Blackman, Y. Shi, S. Klein, M. Cernaianu, D. Doira, P. Genuche, A. Arefiev, "Electron acceleration from transparent targets irradiated by ultra-intense helical laser beams," Commun. Phys. **5**, 116, (2022).
43. L. Allen and M. Padgett, "The Poynting vector in Laguerre–Gaussian beams and the interpretation of their angular momentum density," Opt. Commun. **184**, 1-4, (2000).
44. X. Zhong, Y. Zhao, G. Ren, S. He, and Z. Wu, "Influence of finite apertures on orthogonality and completeness of Laguerre-Gaussian beams," IEEE, **6**, (2018).
45. V. Arrizon, U. Ruiz, R. Carrada, and L. A. Gonzalez, "Pixelated phase computer holograms for the accurate encoding of scalar complex fields," J. Opt. Soc. A. **24**, 11, (2007).
46. E. Bolduc, N. Bent, E. Santamato, E. Karimi, and R. W. Boyd, "Exact solution to simultaneous intensity and phase encryption with a single phase-only hologram," Opt. Lett. **38**, 18, (2013).
47. J. Verbeek, H. Tian, and P. Schattschneider, "Production and application of electron vortex beams," Nature, **467**, 301-304, (2010).
48. M. Uchida and A. Tonomura, "Generation of electron beams carrying orbital angular momentum," Nature, **464**, 737-739, (2010).
49. V. Grillo, G. C. Gazzadi, E. Mafakheri, S. Frabboni, E. Karimi, and R. W. Boyd, "Holographic generation of highly twisted electron beams," Phys. Rev. Lett. **114**, 034801, (2015).